\documentclass[11pt]{article} 
\usepackage[ a4paper,
 total={170mm,257mm},
 left=20mm,
 top=20mm]{geometry}
\usepackage[normalem]{ulem}
\usepackage{graphicx}
\usepackage{hyperref}
\usepackage{amssymb}
\usepackage{authblk}
\usepackage{lineno}
\usepackage{color}

\date{\today}

\title{ANTARES upper limits on the multi-TeV neutrino emission from the GRBs detected by IACTs}

\author[1,2]{A.~Albert}
\author[3]{M.~Andr\'e}
\author[4]{M.~Anghinolfi}
\author[5]{G.~Anton}
\author[6]{M.~Ardid}
\author[7]{J.-J.~Aubert}
\author[8]{J.~Aublin}
\author[8]{B.~Baret}
\author[9]{S.~Basa}
\author[10]{B.~Belhorma}
% ...
\author[  ]{                                                                                                                             (The ANTARES Collaboration)}

\affil[1]{\scriptsize{Universit\'e de Strasbourg, CNRS,  IPHC UMR 7178, F-67000 Strasbourg, France}  }  
\affil[2]{\scriptsize{Universit\'e de Haute Alsace, F-68200 Mulhouse, France}}
\affil[3]{\scriptsize{Technical University of Catalonia, Laboratory of Applied Bioacoustics, Rambla Exposici\'o, 08800 Vilanova i la Geltr\'u, Barcelona, Spain}}
\affil[4]{\scriptsize{INFN - Sezione di Genova, Via Dodecaneso 33, 16146 Genova, Italy}}
\affil[5]{\scriptsize{Friedrich-Alexander-Universit\"at Erlangen-N\"urnberg, Erlangen Centre for Astroparticle Physics, Erwin-Rommel-Str. 1, 91058 Erlangen, Germany}}
\affil[6]{\scriptsize{Institut d'Investigaci\'o per a la Gesti\'o Integrada de les Zones Costaneres (IGIC) - Universitat Polit\`ecnica de Val\`encia. C/  Paranimf 1, 46730 Gandia, Spain}}
\affil[7]{\scriptsize{Aix Marseille Univ, CNRS/IN2P3, CPPM, Marseille, France}}
\affil[8]{\scriptsize{Universit\'e de Paris, CNRS, Astroparticule et Cosmologie, F-75006 Paris, France}}
\affil[9]{\scriptsize{Aix Marseille Univ, CNRS, CNES, LAM, Marseille, France }}
\affil[10]{\scriptsize{National Center for Energy Sciences and Nuclear Techniques, B.P.1382, R. P.10001 Rabat, Morocco}}
% ...

\begin{document} 


\maketitle 


\begin{abstract}
The first gamma-ray burst detections by Imaging Atmospheric Cherenkov Telescopes have been recently announced: GRB 190114C, detected by MAGIC, GRB 180720B and GRB 190829A, observed by H.E.S.S. A dedicated search for neutrinos in space and time coincidence with the  gamma-ray emission observed by IACTs has been performed using ANTARES data. The search covers both the prompt and afterglow phases, yielding no neutrinos in coincidence with the three GRBs studied. Upper limits on the energetics of the neutrino emission are inferred. The resulting upper limits are several orders of magnitude above the observed gamma-ray emission, and they do not allow to constrain the available models.
\end{abstract}

\flushbottom


\section{\label{s:intro} Introduction}

Gamma-ray bursts (GRBs) are one of the most energetic explosive phenomena observed in the Universe through electromagnetic (EM) radiation. However, the physical mechanisms at play during these transient phenomena are not yet fully understood. In particular, if hadronic acceleration takes place within the released outflow, neutrinos and cosmic rays might be produced with the GRB~\cite{waxman,waxman2}. The observation of astrophysical neutrinos in coincidence with the detected EM radiation would confirm hadronic acceleration and help to improve the understanding of the underlying processes at work.

Two different phases can be distinguished in the EM light-curves observed: a prompt emission extending up to a few tens of seconds where the gamma-ray radiation (believed to arise from internal shocks) presents the highest luminosity, and a long lasting afterglow component, dominated by X-ray and radio emissions~\cite{GRBreview}.

Large ground-based Cherenkov detectors observing the showers induced in the atmosphere by high-energy photons (or charged cosmic rays) are well-suited for the observation of GRB gamma rays at $\mathcal{O}$(TeV) energies. Current instruments capable of detecting the gamma-ray emission up to these very-high energies (VHE) are the three Imaging Atmospheric Cherenkov Telescopes (IACTs), which perform stereoscopic observations: H.E.S.S., MAGIC and VERITAS. Also the HAWC and LHAASO observatories, which consist of particle detectors observing the secondary particles from the showers initiated in the atmosphere, are capable to detect and distinguish primary gamma-rays from the charged cosmic rays in the same energy range. 

Three GRBs have lead to the detection of photons by IACTs: GRB 180720B observed by H.E.S.S.~\cite{HESS1}, GRB 190114C detected by MAGIC~\cite{MAGIC1,MAGIC2} and GRB 190829A~\cite{GCN_190829_HESS} (ATel \#13052), also revealed by the H.E.S.S. telescope. The main general features of the three GRBs are summarised in table~\ref{t:grbs}. In particular, the trigger time of the event ($T_0$), the time during which 90\% of the total gamma-ray luminosity is released ($T_{90}$), which is given relative to $T_0$, the position of the source in equatorial coordinates and its redshift ($z$) are provided.

Interestingly, the observation of gamma rays with energies $>$100~GeV taking place hundreds to thousands of seconds after the GRB explosion, namely during the afterglow phase, violates the maximum photon synchrotron energy limit~\cite{Fermi_sync}. In fact, the synchrotron radiation by electrons is characterized by a maximum energy above which this process is not enough to explain the observed photons. This happens when energy losses, which become relevant in the afterglow phase, overcome the efficiency of the acceleration process that yields the synchrotron emission. The violation of the synchrotron limit indicates that an additional mechanism may take place other than synchrotron radiation, opening the possibility for the presence of an hadronic contribution needed to model the full light-curve, motivating a neutrino follow-up. 

Even though the VHE photons observed from these three GRB detections by IACTs can be explained by leptonic models within the synchrotron self-Compton scenario~\cite{syncmodel}, a subdominant hadronic component is not ruled out by current data, and these sources remain of interest as possible cosmic neutrino emitters. In fact, photo-hadronic interactions (p$-\gamma$) are expected to contribute to the GRB emission, both in the prompt phase~\cite{waxman,waxman2,neucosm} as well as in the afterglow~\cite{GRB_afterglow1,GRB_afterglow2}. The latter case was recently explored in~\cite{GRBhadronic}, where the authors show that this mechanism is capable of reproducing the $\mathcal{O}$(TeV) observations of GRB 190114C and GRB 180720B.

Previous searches have not detected a significant neutrino flux associated with standard GRBs, namely emitting in the keV to GeV domain~\cite{brightGRBs,GRBsIC,Angela}. In these papers, a model-dependent search for a prompt emission seen as upgoing muon neutrinos in the detector is carried out, covering from the event trigger at $T_0$ up to $T_{90}$. However, GRBs emitting in the TeV domain might possibly behave differently from those previously investigated. Indeed, the presence of a high-energy component in their radiation spectra might constitute a promising feature also in terms of neutrino emission.

In this paper, a dedicated study is carried out for each of the three promising GRBs to investigate a potential hadronic component that could be present in the EM data. The document is organised as follows: the ANTARES telescope and principles of neutrino searches are introduced in section~\ref{s:antares}. The analysis method and the event selection are explained in section~\ref{ss:selection}. Sections \ref{s:GRB180720}, \ref{s:GRB190114} and \ref{s:GRB190829} focus on each of the three GRB events, describing their characteristics and providing details and results of the neutrino search performed. The conclusions are drawn in section~\ref{s:conc}.

\begin{table}[!h]
    \caption{General features of the three GRB events considered in the analysis. The first column is the GRB event name. The next two columns indicate the date and time ($T_0$) of the event trigger. The fourth column provides the $T_{90}$, which is measured with respect to $T_0$. The last two columns contain the redshift and position of the source in equatorial coordinates.}\vspace{0.2cm}
    \label{t:grbs}

    \centering
    \resizebox{0.9\textwidth}{!}{%
    \begin{tabular}{|c|c|c|c|c|c|}\hline
        { Event} & Date & {$T_0$ (UTC time)} & $T_{90}$ (s) & $z$ & (RA; $\delta$) \\ \hline
    GRB 180720B & 20/07/2018 & 14h:21m:40s & 50 & 0.653 & (0.53$^{\circ}$; $-$2.93$^{\circ}$) \\ \hline
    GRB 190114C & 14/01/2019 & 20h:57m:03s & 116 & 0.42 & ($-$54.51$^{\circ}$; $-$26.94$^{\circ}$) \\ \hline
    GRB 190829A & 29/08/2019 & 19h:55m:53s & 63 &  0.0785 & (45.6$^{\circ}$; $-$7.1$^{\circ}$) \\ \hline 
    \end{tabular}}
\end{table}

\section{ANTARES neutrino search}
\label{s:antares}

The ANTARES telescope~\cite{ANTARES}, located in the Mediterranean Sea, is the longest-lived neutrino detector operating in the Northern Hemisphere. The detector consists of a three-dimensional array of 885 optical modules (OMs) distributed over 12 vertical strings. The lowest OM of each detection string is located at a depth of about 2375~m. Each OM houses a 10” photomultiplier tube. OMs are arranged in groups of three to form a storey. 

In astrophysical neutrino searches with neutrino telescopes, the atmospheric muon background component is dominant. For this reason, the standard astrophysical neutrino analyses select events seen as upgoing in the detector, since only neutrinos can travel through the Earth~\cite{PS2012}. This defines the horizon of neutrino telescopes. However, with an astronomical signal defining the search sky position and time window, the background is significantly reduced, and it is possible to extend the search over the full sky~\cite{GWHENO2}. As a consequence, both hemispheres (sky regions above and below the horizon), with different background conditions, are considered in the analysis.

Two different event topologies can be identified in neutrino telescopes: tracks and showers. On the one hand, muons produced in the interaction of cosmic rays in the atmosphere, and as a product of muon neutrino charged current interactions in sea water, give rise to long tracks of detectable Cherenkov light. On the other hand, showers induce quasi-spherical light depositions in the detector, which are the result of neutral current interactions of all neutrino flavors, as well as from electron neutrino charged current interactions. The ANTARES detector achieves a median angular resolution of 0.4$^{\circ}$ at 10~TeV for track events~\cite{PS2012}, and about 3$^{\circ}$ for shower events~\cite{TANTRA}. An all-flavor neutrino search is possible using the two event topologies. The very different expected background levels for upward- and downward- going events require different selection criteria, yielding four data samples analysed separately when the two topologies (tracks and showers) are considered.


When an astrophysical transient event, potentially a neutrino source, triggers an alert, a real-time neutrino follow-up is performed to look for candidates in the data. This online search only applies to upgoing track-like events, and it is useful in order to provide rapid information to the community in form of ATELs and GCN circulars. Afterwards, a refined analysis is carried out which incorporates dedicated offline calibrations~\cite{calib1,calib2,calib3} to the data as well as a run-by-run driven Monte Carlo simulation~\cite{ANTMC} for a more precise estimate of the detector response to the searched astrophysical flux.

\section{Analysis method}
\label{ss:selection}

For each of the GRBs analysed in this paper, the time window of the search is adjusted to cover the interval ranging from the trigger time up to the end of the observations by IACT's. In fact, the exact time where neutrinos are expected after the burst and with respect to the $\mathcal{O}$(TeV) gamma rays is highly model dependent~\cite{GRBt1,GRBt2}. Moreover, the observational conditions of IACTs are driven by background light contamination and zenith angle constraints. Therefore, the time coverage of the observations in their sensitive energy range, with respect to the burst onset and the end of the afterglow observations, might just come from their limited observational time with good data conditions, which may result in missing the prompt radiation as well as the very late afterglow emission. Nevertheless, in order to be as model independent as possible, the time window is restricted in this work to the intervals encompassing the detection of the burst and the observation of VHE photons by IACTs. These short time windows present the advantage to allow for a follow-up also when the sources are above the ANTARES horizon (producing downgoing events in the detector), which was the case during the IACTs observations, since the short exposure reduces the overall background. The duration of the time windows ($\delta t_{\rm total}$) that will be used for each of the GRB events, as well as the interval during which the source was seen respectively as upgoing and downgoing ($\delta t_{\rm up}$ and $\delta t_{\rm down}$) are summarised in table~\ref{t:tdur}.

\begin{table}[!h]
    \caption{Duration of the search time windows used in the neutrino follow-up analysis of each GRB event. The hyphen indicates that the source was not below the ANTARES horizon during the search time window.}\vspace{0.2cm}
    \label{t:tdur}

    \centering
    \resizebox{0.46\textwidth}{!}{%
    \begin{tabular}{|c|c|c|c|}\hline
        { Event} & {$\delta t_{\rm total}$} & {$\delta t_{\rm up}$} & {$\delta t_{\rm down}$} \\ \hline
    GRB 180720B & 12.1~h & 7.7~h & 4.4~h \\ \hline
    GRB 190114C & 2804~s & $-$ & 2804~s \\ \hline
    GRB 190829A & 8.1~h & 2.85~h & 5.25~h \\ \hline 
    \end{tabular}}
\end{table}


The event selection applied in this work is the one commonly followed in ANTARES for transient sources, detailed in~\cite{GWHENO2}. The spatial search region around each source, also called the region of interest, is optimised for each event. The quality cuts of the reconstruction that allow to reach the most constraining limits are optimised by maximising the signal expectation assuming an $E^{-2}$ neutrino spectrum while keeping the false alarm rate below the desired threshold. The $E^{-2}$ spectrum is expected in the case of the generic Fermi acceleration mechanism~\cite{fermiacc}. It is the sole assumption made for the signal expectation in the analysis. 

The selection criteria is chosen so that one neutrino candidate event passing the analysis cuts found in time and space coincidence with the gamma-ray emission leads to a 3$\sigma$ detection. Using Poisson statistics, this requirement is equivalent to imposing that the probability to have one background event is smaller than $p_{3\sigma}$ = $2.7 \times 10^{-3}$. The quality parameters used to optimise each event sample are summarised in table~\ref{t:cuts}. 

\begin{table}[!h]
    \caption{Quality parameters of the reconstruction used for the event selection, for each event sample.}\vspace{0.2cm}
    \label{t:cuts}

    \centering
    \resizebox{0.92\textwidth}{!}{%
    \begin{tabular}{|c|c|}\hline
        {\bf Sample} & {\bf Quality parameters} \\ \hline
        Upgoing tracks & Likelihood of the track reconstruction ($\Lambda$)~\cite{PS2012} \\ \hline
        Downgoing tracks & $\Lambda$ and track energy estimate ($N_{\rm hits, tr}$)~\cite{PS2012} \\ \hline
        Upgoing showers & Likelihood ratio ($L_{\mu}$) \cite{TANTRA} and Random Decision Forest classifier \cite{DUSJ} \\ \hline
        Downgoing showers & $L_{\mu}$ and shower energy estimate ($N_{\rm hits, sh}$)~\cite{TANTRA} \\ \hline
    \end{tabular}}
\end{table}


In case no neutrino is found in time and space coincidence, upper limits on the neutrino spectral fluence, the total energy release in neutrinos, and the fraction of energy going into pions over that going into electrons, will be derived. The systematic uncertainties related to the ANTARES detector, detailed in ref.~\cite{GWHENO2}, have been studied. The evaluated total systematic uncertainty on the neutrino fluence upper limit is of about 33\% for upgoing and 42\% for downgoing events.

\section{\label{s:GRB180720} GRB 180720B and the gamma-ray observation by H.E.S.S.}

On 20th July 2018, the GBM instrument on board the Fermi satellite triggered on the gamma-ray emission from GRB 180720B at UTC time $T_0 = 14$h$:21$m$:40$s~\cite{GCN_180720_Fermi}. The burst was also triggered on by Swift-BAT, 5~s later~\cite{GCN_180720_Swift}. This event appears to be the 6th brightest GRB ever observed by Fermi-GBM, and the GRB with the second highest energy flux observed by Swift-XRT~\cite{GCN_180720_Swift} in the afterglow (11~h after $T_0$), with a very long X-ray plateau. The source has been measured by the VLT to be at a redshift of $z$ = 0.653, with a $T_{90}$ = 48.9~s measured by Fermi-GBM~\cite{HESS1}. The source was positioned at equatorial coordinates (RA;~$\delta$)~=~(0.53$^{\circ}$;~$-$2.93$^{\circ}$).

The source GRB 180720B entered in the H.E.S.S. field of view about 10.1~h after $T_0$. An observation was performed with a total exposure of 2~h that led to a 5$\sigma$ detection of $\mathcal{O}$(TeV) gamma-ray emission~\cite{HESS1}. The H.E.S.S. observation was not notified to the community, preventing a fast follow-up by ANTARES.


\subsection{\label{ss:method1} Neutrino follow-up}

The search for neutrino emission from the GRB starts 350~s before $T_0$ to include potential precursors~\cite{baret}, and extends 10.1~h (until the start of the H.E.S.S. observation) plus 2~h (to take into account the H.E.S.S. exposure) after the $T_0$. During the search window, [$-$350~s, +12.1~h], the position of the source changes in the ANTARES local detector frame. As mentioned in section~\ref{ss:selection}, the two sky hemispheres are considered separately due to the different background conditions. This corresponds to a 7.7~h search below the ANTARES horizon (i.e., for events seen as upgoing in the detector local frame), including $T_0$ and the prompt emission, and a 4.4~h search for downgoing events (above the horizon), covering the H.E.S.S. observation of the afterglow.

The size of the region of interest and the quality cuts are optimised following the method described in section~\ref{ss:selection}. For tracks, the optimal search region results in a circle of 2$^{\circ}$ radius around the source for both upgoing and downgoing events. The optimised region of interest for the shower search is a circle of radius 24$^{\circ}$ and 7$^{\circ}$, for upgoing and downgoing events, respectively. The higher background rates in the downgoing sky due to the atmospheric muon flux explains the reduced angular search window with respect to upgoing events. The analysis yields no neutrino counterpart to the GRB signal after unblinding the data.



\subsection{\label{ss:results1} Upper limits on the neutrino emission}
The time-integrated neutrino flux from a given astrophysical source ($dN/dE_{\nu}$, in GeV$^{-1} \cdot$ cm$^{-2}$) is used to compute the expected number of neutrinos that would be observed from a source at declination $\delta$ in a detector with effective area $ A_{eff}(E_{\nu},\delta)$:
\begin{equation}
    N_{\nu} = \int \frac{dN}{dE_{\nu}}(E_{\nu}) A_{eff}(E_{\nu},\delta) dE_{\nu}.
\end{equation}

Assuming that the neutrino differential energy distribution is a power law with spectral index $\gamma = 2$, then the spectral fluence at the detector can be expressed as follows: 
\begin{equation}
    \phi_0 = E_{\nu}^{2} \frac{dN}{dE_{\nu}} {\rm [GeV \cdot cm^{-2}]} 
    \label{eq:flu}
\end{equation}
Fluence upper limits (ULs) can easily be derived from Eq.~\ref{eq:flu}, considering the Poisson 90\% confidence level upper limit when no event is observed, $N_{\nu}^{90 \%}$= 2.3 events. This leads to an upper limit on the neutrino spectral fluence: $\phi_{0, \rm up}^{90\%}$ = 1.5 GeV$\cdot$cm$^{-2}$ for the upgoing search, and $\phi_{0, \rm down}^{90\%}$ = 10 GeV$\cdot$cm$^{-2}$ for the downgoing search. Since the redshift of the source is known, the luminosity distance $D_L (z)$ can be computed so that the fluence upper limits can be converted into upper limits on the total isotropic neutrino energy emitted within the 5$-$95\% energy range of the analysis (reported in table~\ref{t:flulim}) following:

\begin{equation}
E_{\nu,{\rm iso} } = \frac{ 4 \pi D_L(z)^{2} }{ 1+z } \int^{E^{95\%}}_{E^{5\%}} E_{\nu}^{-2} \phi_0^{90\%} E_{\nu} dE_{\nu}.
\label{eq:limit}
\end{equation}  

The cosmological parameters from~\cite{Hubble} are used to obtain the luminosity distance from the measured redshift. The results obtained are $E_{\nu,{\rm iso} }^{\rm up} \lesssim$ 2$\times 10^{55}$~erg [2.5~TeV; 4.0~PeV] and and $E_{\nu,{\rm iso} }^{\rm down} \lesssim$ 1$\times 10^{56}$~erg [20~TeV; 30~PeV] for upgoing and downgoing events, respectively. These neutrino limits are about 30 (upgoing) and 200 (downgoing) times above the isotropic energy inferred from EM observations, $E_{\gamma, {\rm iso}}$ = 6$\times 10^{53}$~erg [50; 300~keV]~\cite{HESS1}.

Moreover, one can make use of Eq. (5) in~\cite{Yacobi} to set constrains on the fraction of energy going respectively into pions and electrons:
\begin{equation}
F_{\nu} = \frac{1}{12} \frac{f_{\pi}}{f_{e}} \frac{F_{\gamma}}{{\rm ln}(E_{\rm max,e}/E_{\rm min,e})},
\label{eq:fraction}
\end{equation}
where $E_{\rm max, e}$ and $E_{\rm min, e}$ represent respectively the maximum and minimum energy of electrons radiating photons with fluence $F_{\gamma}$. The factor 1/12 comes, first from the fact that the fraction of the proton energy that goes into neutrinos is $\sim f_{\pi}$/4, $f_{\pi}$ being the fraction that goes into pions, and second because we are computing a one-flavor neutrino limit (1/3). Now, we are interested in obtaining this same Eq.~(\ref{eq:fraction}) as a function of the photon and neutrino energies only. By relating the electron's energy to the emitted photon's energy as $E_\gamma \propto E_e^2$ (holding e.g. in the case of synchrotron and inverse Compton emission), and by moving from fluence to isotropic energy, analogously to Eq.~(\ref{eq:limit}), we derive:
\begin{equation}
\label{eq:yacobi}
\frac{f_{\pi}}{f_{e}} = 12 \frac{1}{2} {\rm ln} \left( \frac{E_{\rm max, \gamma}}{E_{\rm min, \gamma}} \right) \frac{E_{\rm iso, \nu}}{E_{\rm iso, \gamma}} \frac{{\rm ln}(E_{\rm max, \gamma}/E_{\rm min, \gamma})}{{\rm ln}(E_{\rm max, \nu}/E_{\rm min, \nu})} ,
\end{equation}
where $E_{\rm max, \gamma}$ and $E_{\rm min, \gamma}$ are the maximum and minimum energy of the range of the EM measurement of $E_{\rm iso, \gamma}$. $E_{\rm min, \nu}$ and $E_{\rm max, \nu}$ are defined by the 5$-$95\% energy range of the neutrino analysis. All these values are reported in table~\ref{t:flulim}. The resulting limits are $(f_{\pi}/f_{e})_{\rm up} \lesssim$80 and  $(f_{\pi}/f_{e})_{\rm down} \lesssim$600, respectively for upgoing and downgoing events.
\section{\label{s:GRB190114} GRB 190114C and the observation by MAGIC}

On January 14th 2019, the BAT instrument on board the Swift satellite triggered on the gamma-ray emission from GRB 190114C at $T_0 = 20$h$:57$m$:03$s UTC~\cite{GCN_190114_Swift}. The burst has also been detected by other satellite-based instruments like Fermi-GBM~\cite{GCN_190114_Fermi} and Integral/SPI-ACS~\cite{GCN_190114_integral}. The MAGIC telescope performed an observation of the source during about 40 minutes, starting $\sim$ 50~s after the trigger. MAGIC observations led to a $>$20$\sigma$ detection~\cite{GCN_190114_MAGIC}. The source has been estimated to be at a redshift of $z = 0.42$~\cite{GCN_190114_redshift}. The $T_{90}$ has been measured to be about 116~s by Fermi-GBM [50~keV; 300~keV]~\cite{GCN_190114_Fermi} and about 362~s by Swift-BAT [15~keV; 350~keV]~\cite{GCN_190114_BAT}. The GRB was located at equatorial coordinates (RA; $\delta$) = ($-$54.51$^{\circ}$; $-$26.94$^{\circ}$).

For this event, the total isotropic energy observed by Fermi-GBM during $T_{90}$ was $E_{\gamma,{\rm iso} }$ 2.5$\times 10^{53}$~erg in the [1 keV; 10 MeV] interval~\cite{MAGIC1}. The value measured by MAGIC in the [300~GeV; 1~TeV] energy band was $E_{\gamma,{\rm iso} }$ = 2$\times 10^{52}$~erg~\cite{MAGIC2}.

\subsection{\label{ss:method2} Neutrino follow-up}

At the time of the alert, the source was above the ANTARES horizon, which prevented a real-time follow-up. A dedicated search over the ANTARES dowgoing sky using tailored offline calibrations and reconstruction of the data is carried out. Neutrino emission is searched over a time window of 2804~s, in the range [$T_0-$350, $T_0$+2454]~s. By starting 350~s before the trigger time ($T_0$) the search takes into account possible GRB precursors, as motivated in~\cite{baret}, and by extending it up to 2454~s after $T_0$, it is covering the entire prompt burst phase up to the end of the observation by MAGIC in the afterglow~\cite{MAGIC2}.

For each sample, the search region is optimised following the same procedure and assumptions as the ones described in section~\ref{ss:selection}. For tracks, the good angular resolution leads to an optimised region for the search of 2$^{\circ}$ radius around the source position. The less accurate direction estimate for showers leads to a search region of 22$^{\circ}$ around the source. After the selection is applied and the data unblinded, no neutrino is found in ANTARES data in time and space coincidence with the electromagnetic GRB emission. 

\subsection{\label{ss:results2} Upper limits on the neutrino emission}

From the non-detection, an upper limit on the neutrino spectral fluence can be derived to constrain the neutrino emission from this source. 


The ANTARES 90\% UL is computed assuming an $E^{-2}$ neutrino spectrum over an extended window of 2804~s. The obtained integrated neutrino fluence UL is $\phi_0^{90 \%}$ = 1.6~GeV$\cdot$cm$^{-2}$. For comparison, the results by the IceCube neutrino follow-up provides a limit of $\phi_0^{90 \%}$ = 0.44~GeV$\cdot$cm$^{-2}$ (ATel \#12395).

\begin{figure}[!h]
    
    \centering
    \includegraphics[scale=0.57]{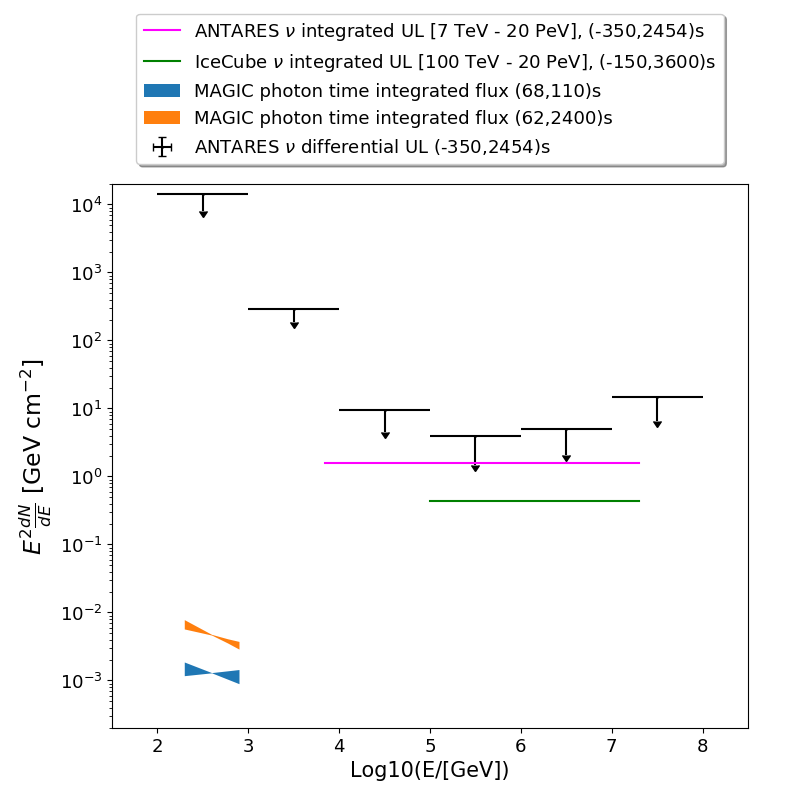}
    
    \caption{ANTARES 90\% differential (black arrows) and integrated (pink line) spectral fluence upper limits as a function of the neutrino energy  for GRB 190114C. The pink line shows the limit integrated in the 5$-$95\% energy range of the analysis (see table~\ref{t:flulim}). An extended emission over 2804~s, with the time interval detailed in the legend, is used in the analysis, and an E$^{-2}$ neutrino spectrum is assumed. The IceCube upper limits (green line) from ATel \#12395 is shown for comparison. Also shown is the MAGIC gamma-ray spectral fluence for the first time bin of the analysis (68-110~s, blue band), and for the overall time window (62-2400~s, orange band).}
    \label{fig:limitmagic}
\end{figure}

Results are shown in figure~\ref{fig:limitmagic} in the form of a differential fluence upper limit as a function of the energy. MAGIC data in~\cite{MAGIC2} (Extended Data Table 1) are used to infer the differential gamma-ray fluence corresponding to the MAGIC flux over the time window $\Delta t$ as follows: and integrating the MAGIC flux normalisation ($\phi_0^{\gamma}$) over the selected time window ($\Delta t$),
\begin{equation}
    E_{\gamma}^{2} \frac{dN}{dE_{\gamma}} = \phi_0^{\gamma} \times \left( \frac{E_{\gamma}}{E_{\rm pivot}} \right) ^{- \alpha} \times \Delta t \times E_{\gamma}^{2} ,  
\end{equation}
where $E_{\rm pivot}$, $\phi_0^{\gamma}$ and $\alpha$ are respectively the pivot energy, the flux normalisation and photon index describing the spectrum as fitted by MAGIC. $E_{\gamma}$ is the photon energy in the range of the MAGIC observation [300~GeV$-$1~TeV]~\cite{MAGIC1,MAGIC2}. 

Such limits are derived for both the first data analysis time interval of the MAGIC observation (68-110~s from the Swift trigger), corresponding to the blue band in figure~\ref{fig:limitmagic}, and over the full time window adopted for the spectral analysis (62-2400~s), corresponding to the orange band in figure~\ref{fig:limitmagic}. Both are compared, including the corresponding systematic uncertainties in the spectral fit, to the ANTARES and IceCube neutrino upper limits. The neutrino limit obtained in this work is orders of magnitude above the MAGIC flux. Therefore, the hadronic content of the source cannot be constrained.

As in section~\ref{ss:results1}, if the distance of the source is known, one can evaluate the upper limit on the total energy emitted in neutrinos within the 5$-$95\% energy range of the search using Eq.~\ref{eq:limit}. For this event, the result obtained is $E_{\nu,{\rm iso} } \lesssim$ 8$\times 10^{54}$~erg [7~TeV; 20~PeV], which is 400 times above the one measured by the MAGIC telescope ($E_{\gamma,{\rm iso} }$ = 2$\times 10^{52}$~erg). 

Finally, limits on the pion to electron energy fraction can also be derived, yielding as a result $(f_{\pi}/f_{e})_{\rm down} \lesssim 2 \times 10^{3}$, computed considering the Fermi-GBM measurement: $E_{\gamma,{\rm iso} }$ = 2.5$\times 10^{53}$~erg in the interval [1 keV; 10 MeV].


\section{\label{s:GRB190829} GRB 190829A and the observation by H.E.S.S.}

The GBM instrument on board the Fermi satellite triggered on the gamma-ray emission from GRB 190829A on 29th August 2019 at the time $T_0 = 19$h$:55$m$:53$s UTC~\cite{GCN_190829_Fermi}. The source has been measured to be at a redshift of $z = 0.0785$~\cite{GTC,GCN_190829_redshift}, with a $T_{90}$ of 63~s~\cite{GCN_190829_GBM}. The observation of this source with the H.E.S.S. telescope started 4~h and 20~min after $T_0$, and the total exposure of 3~h and 40~min led to a significant ($>$5$\sigma$ excess) detection, with $\sim$ TeV gamma-ray emission observed during the afterglow~\cite{GCN_190829_HESS}. The position of the source is (RA; $\delta$) = (45.6$^{\circ}$; $-$7.1$^{\circ}$). 


The light-curve of GRB 190829A measured by Fermi~\cite{GCN_190829_Fermi} shows two peaks, and the total isotropic energy derived using the best fit to the multi-wavelength spectrum is $E_{\gamma, {\rm iso}} = 3 \times 10^{50}$~erg [1 keV; 10 MeV]~\cite{ongrbHESS2}. The synchrotron self-Compton scenario is compatible with both the limits set from the non-observation by Fermi-LAT and the detection at VHE by H.E.S.S.~\cite{ongrbHESS2}. However, the H.E.S.S. analysis is not yet public and these conclusions are only based on Fermi-LAT upper limits and an estimation of the H.E.S.S. sensitivity.

\subsection{\label{ss:method3} Neutrino follow-up}

A real-time search was performed with ANTARES within $\pm$1~h from $T_0$, and in three degrees around the source position. 
No upgoing tracks were observed~\cite{GCN_190829_ANTARES}. In this work, the dedicated offline neutrino search is presented. The analysis focuses on a potential neutrino emission during the $\sim$8.1~h duration of the EM light-curve of the GRB, up to the end of the observations reported by H.E.S.S. in~\cite{GCN_190829_HESS}. As for the previous GRBs, a short window before $T_0$ is considered according to ref.~\cite{baret}. During the $\sim$8.1~h, the source changes its position in the local detector frame. Therefore, two separate searches are carried out for the time windows where the source is below and above the ANTARES horizon. This corresponds to a 2.85~h search below the ANTARES horizon (i.e., for upgoing events), including $T_0$ and the prompt emission, and a 5.25~h search for downgoing events (above the horizon), containing the H.E.S.S. observation of the afterglow.

The search is performed separately for the different samples as described in section~\ref{ss:selection}. For tracks, the optimal search region is found to be a circle of 2$^{\circ}$ radius around the source for the upgoing search and 1$^{\circ}$ for downgoing events. For shower events, the optimised region of interest for the search is a circle with 20$^{\circ}$ radius for upgoing events and 9$^{\circ}$ for the downgoing sky.

The same procedure as before is applied, imposing the expected number of background events to be below $2.7 \times 10^{-3}$ over the entire search window. After unblinding the data, no neutrino is found in correlation with the GRB signal. 

\subsection{\label{ss:results3} Upper limits on the neutrino emission}

As for the GRBs discussed above, the 90\% CL upper limits on time-integrated neutrino flux, the total energy emitted in neutrinos, and the fraction of energy going respectively into pions and electrons, are provided. 

The obtained 90\% CL neutrino fluence upper limits are $\phi_{0}^{90\%}$ = 1.4~GeV$\cdot$cm$^{-2}$ for the upgoing search and 4~GeV$\cdot$cm$^{-2}$ for the downgoing search. Using the measured redshift and the relation in Eq.~\ref{eq:limit}, the fluence upper limits are translated into upper limits on the total isotropic neutrino emission, which are: $E_{\nu,{\rm iso} } \lesssim$ 2$\times 10^{53}$~erg [2.5~TeV; 4.0~PeV] for the search using upgoing events and $E_{\nu,{\rm iso} } \lesssim$ 7$\times 10^{53}$~erg [15~TeV; 25~PeV] when searching for downgoing events. Finally, limits on the fraction $(f_{\pi}/f_{e})$ are set using Eq.~\ref{eq:yacobi}. The results obtained considering the EM observations with $E_{\gamma, {\rm iso}} = 3 \times 10^{50}$~erg in the energy range [1 keV; 10 MeV] are: $(f_{\pi}/f_{e})_{\rm up} \lesssim 5 \times 10^{4}$ and $(f_{\pi}/f_{e})_{\rm down} \lesssim 2 \times 10^{5}$, for the upgoing and downgoing searches. Using the data reported by Fermi-GBM in~\cite{GRB190829_2}, which measured $E_{\gamma, {\rm iso}} = 2 \times 10^{50}$~erg [50; 300 keV], the limits become: $(f_{\pi}/f_{e})_{\rm up} \lesssim 3 \times 10^{3}$ and $(f_{\pi}/f_{e})_{\rm down} \lesssim 1 \times 10^{4}$, for the upgoing and downgoing searches. This comparison shows the dependence on the limit with the energy range coverage for the $E_{\gamma, {\rm iso}}$ measurement.



\begin{table}[!h]
    \caption{Upper limits on the neutrino spectral fluence, the total energy emitted in neutrinos within the 5$-$95\% energy range of the analysis, and the fraction of energy going to pions to that going to electrons, $f_{\pi}/f_e$. Results are presented for the three GRB searches, separately for upgoing and downgoing events. The hyphen indicates that the corresponding GRB was not seen as upgoing during the gamma-ray emission. The systematic uncertainties on $\phi_0^{90\%}$ are reported in section~\ref{ss:selection}. The last row shows the isotropic photon energy measured, $E_{\gamma, {\rm iso}}$. For event GRB 190829A, two $E_{\gamma, {\rm iso}}$ values in different energy ranges are considered to evaluate the $f_{\pi}/f_e$ limits.}\vspace{0.2cm}
    
    \label{t:flulim}
    \centering
    \resizebox{\textwidth}{!}{%
    \begin{tabular}{|c|c|c|c|}\hline
{} & {\bf GRB 180720B} & {\bf GRB 190114C} & {\bf GRB 190829A} \\ \hline
$\delta t_{\rm up}$ & [$T_0-$350~s, $T_0$+7.6~h] & - & [$T_0-$350~s, $T_0$+2.85~h] \\ \hline
$\phi_0^{90\%}$~$_{\rm up}$ (GeV$\cdot$cm$^{-2}$) & $\lesssim$1.5 & - & $\lesssim$1.4 \\ \hline 
$\delta t_{\rm down}$ & [$T_0$+7.6~h, $T_0$+12.1~h] & [$T_0-$350~s, $T_0$+2454~s] & [$T_0$+2.85~h, $T_0$+8.1~h] \\ \hline
$\phi_0^{90\%}$~$_{\rm down}$ (GeV$\cdot$cm$^{-2}$) & $\lesssim$10 & $\lesssim$1.6 & $\lesssim$4 \\ \hline
$E^{\rm up}_{5-95\%}$ & 2.5 TeV $-$ 4.0 PeV & - & 2.5 TeV $-$ 4.0 PeV \\ \hline
$E_{\nu, \rm iso}^{90\%}$ upgoing (erg) & $\lesssim$2$\times 10^{55}$ & - & $\lesssim$2$\times 10^{53}$ \\ \hline
$(\frac{f_{\pi}}{f_{e}})_{\rm up}$ & $\lesssim$80 & - & $\lesssim5 \times 10^{4}$ $-$ $\lesssim3 \times 10^{3}$ \\ \hline
$E^{\rm down}_{5-95\%}$ & 20 TeV $-$ 30 PeV & 7 TeV $-$ 20 PeV & 15 TeV $-$ 25 PeV \\ \hline
$E_{\nu, \rm iso}^{90\%}$ downgoing (erg) & $\lesssim$1$\times 10^{56}$ & $\lesssim$8$\times 10^{54}$ & $\lesssim$7$\times 10^{53}$ \\ \hline
$(\frac{f_{\pi}}{f_{e}})_{\rm down}$ & $\lesssim$600 & $\lesssim2 \times 10^{3}$  & $\lesssim2 \times 10^{5}$ $-$ $\lesssim 1\times 10^{4}$ \\ \hline  
$E_{\gamma, \rm iso}$ (erg) & 6$\times 10^{53}$ [50; 300 keV] & 2.5$\times 10^{53}$ [1 keV; 10 MeV] & 3$\times 10^{50}$ [1 keV; 10 MeV] \\
 &  & 2$\times 10^{52}$ [300 GeV; 1 TeV] & 2$\times 10^{50}$ [50 keV; 300 keV] \\ \hline
    \end{tabular}}
\end{table}

\section{\label{s:conc} Conclusions}
The offline ANTARES neutrino searches presented here yields no neutrinos observed in correlation with the three GRBs detected so far by ground-based gamma-ray facilities (IACTs) at $\mathcal{O}$(TeV) energies. This null result allows to set upper limits on the neutrino spectral fluence and on the isotropic energy radiated through neutrinos within the 5$-$95\% energy range of the search for each GRB event, and to compare these limits to the EM observations. Moreover, limits are also be set on the fraction of energy going respectively into pions and electrons. The results for the three gamma-ray bursts observed at very-high energies are summarised in table~\ref{t:flulim}, where these three limits are provided together with the sensitive energy ranges in neutrinos and gamma-rays. No values are given in the upgoing region for GRB 190114C because the source was downgoing in the ANTARES local detector frame during the entire search time window. The resulting limits on the neutrino fluence and isotropic energy are orders of magnitude above the EM data, and do not allow to constrain any of the available models.


Taking GRB 190114C, the neutrino spectral fluence upper limits for the IceCube and ANTARES detectors can be compared within the 5$-$95\% sensitive energy range of the analysis. The source was seen as downgoing for both detectors during the search period. 

The results on the 90\% one-flavor fluence normalisation are 0.44 GeV$\cdot$cm$^{-2}$ [100~TeV; 20~PeV] (IceCube) and 1.6 GeV$\cdot$cm$^{-2}$ [7~TeV; 20~PeV] (ANTARES). As most GRB model predictions show a strong spectral dependence, it appears extremely important to cover the full energy range. While the IceCube analysis was performed only for muon-neutrinos, the ANTARES search is sensitive to all neutrino flavors. Therefore, the complementarity of IceCube and ANTARES in the different sky regions and energy ranges, as well as in flavor sensitivity, is relevant in this kind of searches. Furthermore, even though the source has a negative declination, i.e. it is seen as upgoing most of the time for ANTARES, it was above the ANTARES horizon during the MAGIC observations. If the GRB had happened at another time, an analysis searching below the ANTARES horizon would have provided better limits by a factor of 4 up to 10.

The present limits provided by ANTARES on individual GRB studies are consistent with standard assumption of energy partition in GRB hadronic scenarios. Moreover, constraints on hadron acceleration efficiency during the prompt phase of GRBs are derived in model dependent searches~\cite{brightGRBs}. The results presented here are comparable to previous ANTARES searches. In addition, stacking studies performed by ANTARES and IceCube have already set stringent constraints on the parameter space available for GRBs being significant contributers to the observed astrophysical diffuse neutrino flux~\cite{GRBsIC,Angela}. 

The different models for the expected neutrino flux from GRBs vary greatly in their predictions, and on model parameters, making it difficult to predict which GRBs are most likely to produce a significant neutrino flux. In fact, part of the most relevant effects on the assumed parameters concerning the neutrino prediction were studied in~\cite{Angela}. In particular, the energy of any neutrinos that may be produced, their delay with respect to the gamma-ray emission~\cite{nugrbs} and, in general, the multi-messenger lightcurves from gamma-ray bursts predicted in the case of the commonly used Internal Shock Model~\cite{MM_GRBs}, are uncertain. Moreover, the observed TeV emission from these GRBs might be due to a late-time hadronic acceleration mechanism which is currently not taken into account by the available models. For these reasons, the authors warn the reader from a hasty interpretation of the comparison of the limits derived from the present work with the models, and also with respect to the EM observations.

The uniqueness of the observed TeV emission from these GRBs, reveiling the presence of an additional (potentially hadronic) acceleration mechanism, suggests that these GRBs may have a neutrino flux associated which is not present in the previously studied GRBs, making them perhaps the best candidates among detected GRBs for a neutrino production. Nonetheless, the experimental outcome is that the ANTARES detector has not observed evidence for a neutrino flux correlated with the three events, and the provided neutrino upper limits are not constraining the presence of an hadronic emission.



\section*{Acknowledgments}
The authors acknowledge the financial support of the funding agencies:
Centre National de la Recherche Scientifique (CNRS), Commissariat \`a
l'\'ener\-gie atomique et aux \'energies alternatives (CEA),
Commission Europ\'eenne (FEDER fund and Marie Curie Program),
Institut Universitaire de France (IUF), LabEx UnivEarthS (ANR-10-LABX-0023 and ANR-18-IDEX-0001),
R\'egion \^Ile-de-France (DIM-ACAV), R\'egion
Alsace (contrat CPER), R\'egion Provence-Alpes-C\^ote d'Azur,
D\'e\-par\-tement du Var and Ville de La
Seyne-sur-Mer, France;
Bundesministerium f\"ur Bildung und Forschung
(BMBF), Germany; 
Istituto Nazionale di Fisica Nucleare (INFN), Italy;
Nederlandse organisatie voor Wetenschappelijk Onderzoek (NWO), the Netherlands;
Council of the President of the Russian Federation for young
scientists and leading scientific schools supporting grants, Russia;
Executive Unit for Financing Higher Education, Research, Development and Innovation (UEFISCDI), Romania;
Ministerio de Ciencia e Innovaci\'{o}n (MCI) and Agencia Estatal de Investigaci\'{o}n:
Programa Estatal de Generaci\'{o}n de Conocimiento (refs. PGC2018-096663-B-C41, -A-C42, -B-C43, -B-C44) (MCI/FEDER), Severo Ochoa Centre of Excellence and MultiDark Consolider, Junta de Andaluc\'{i}a (ref. SOMM17/6104/UGR and A-FQM-053-UGR18), 
Generalitat Valenciana: Grisol\'{i}a (ref. GRISOLIA/2018/119), Spain; 
Ministry of Higher Education, Scientific Research and Professional Training, Morocco.
We also acknowledge the technical support of Ifremer, AIM and Foselev Marine
for the sea operation and the CC-IN2P3 for the computing facilities.





\end{document}